\documentclass[conference]{IEEEtran}
\IEEEoverridecommandlockouts

\usepackage{cite}
\usepackage{amsmath,amssymb,amsfonts}
\usepackage{algorithm}
\usepackage{algorithmic}
\usepackage{graphicx}
\usepackage{booktabs}
\usepackage{tabularx}
\usepackage{array}
\usepackage{multirow}
\usepackage{textcomp}
\usepackage{pifont}
\usepackage{xcolor}
\usepackage{microtype}
\usepackage{etoolbox}

\graphicspath{{figures/}}

\newcommand{\cosmac}{\texttt{CosMac}}

\begin{document}

\title{Network-Guided Automated Synthesis of MAC Units for Neural Network Inference}

\author{%
\IEEEauthorblockN{Zhouyang Wu, Hongyang Pan, Chuyu Wang, Ji Zhuang, 
Keren Zhu\IEEEauthorrefmark{1}, Xin Zhao, and Xuan Zeng\IEEEauthorrefmark{1}}
\IEEEauthorblockA{Fudan University, Shanghai, China%
\thanks{\IEEEauthorrefmark{1}Corresponding authors: Keren Zhu (krzhu@fudan.edu.cn) and Xuan Zeng (xzeng@fudan.edu.cn).}}
}

\maketitle
\suppressfloats[t]

\begin{abstract}
In neural inference accelerators, multiply--accumulate (MAC) datapaths account for the majority of circuit area and critical-path delay, where operator-level numerical formats simultaneously dictate hardware overhead and quantization error. Yet no straightforward mapping exists between hardware cost and end-to-end network accuracy: logic synthesis transforms datapaths in a highly nonlinear manner, while numerical errors propagate and couple across layers.
To address this disconnect, we present \textbf{NAMSI}, an automated \underline{\textbf{n}}etwork-\underline{\textbf{a}}ware \underline{\textbf{M}}AC \underline{\textbf{s}}ynthesis framework tailored for neural \underline{\textbf{i}}nference.
NAMSI characterizes heterogeneous data formats using a format--distribution representation (FD) and leverages topology-aware accuracy and hardware-cost modeling (TaH)---calibrated with actual post-synthesis Area--Delay Product (ADP) measurements---to jointly estimate coupled full-network accuracy and circuit cost.
Furthermore, reversible lattice search (RLS) is introduced to tractably navigate the exponentially scaling design space.
Evaluations across five neural networks on ImageNet-1K demonstrate that under post-training quantization (PTQ), NAMSI achieves $54.2\%$--$60.4\%$ reductions in post-synthesis MAC ADP compared with FP32 while retaining at least $99\%$ of full-precision Top-1 accuracy, surpassing baseline approaches in both hardware efficiency and accuracy preservation.
\end{abstract}


\section{Introduction}
\label{sec:introduction}

Low-precision multiply--accumulate (MAC) computation is central to efficient neural inference, with existing efforts advancing along two main lines. One line optimizes quantization for pre-existing hardware through training, calibration, and reconstruction~\cite{jacob2018quantization,esser2020lsq,nagel2020adaround,li2021brecq,frantar2023gptq,xiao2023smoothquant,lin2024awq}, as well as resource-constrained mixed-precision allocation~\cite{wang2019haq,yao2021hawqv3,cai2020edmips,dotzel2024fliqs,zhang2024mofq}, while treating the underlying arithmetic circuits as fixed. The other line develops low-precision numerical representations~\cite{kuzmin2022fp8,liu2023llmfp4,li2020apot,dotzel2024sf4,tambe2020adaptivfloat,rouhani2020msfp,rouhani2023microexponents,guo2022ant,jang2025blockdialect} alongside specialized datapaths, inference kernels, and MAC architectures~\cite{judd2016stripes,sharma2018bitfusion,xia2024quantllm,crafton2025pareto} for hardware and system efficiency gains. Yet across both lines, end-to-end network post-training quantization (PTQ) accuracy and MAC synthesis cost remain outside a single optimization loop.

\begin{figure}[!t]
    \centering
    \includegraphics[width=0.95\columnwidth]{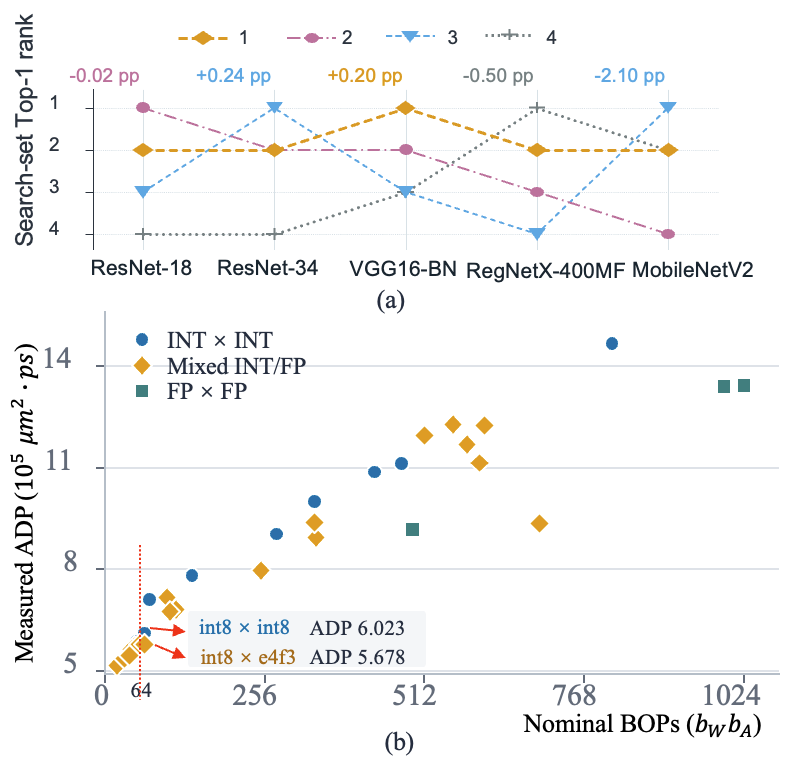}
    \caption{(a) Top-1 accuracy rankings of different quantization schemes across networks. Percentage points (pp) denote the difference between the top-ranked scheme and the corresponding FP32 Top-1 baseline on the same search set. (b) Measured post-synthesis MAC ADP versus nominal BOPs ($b_W b_A$) under a uniform clock period.}
    \label{fig:motivation}
\end{figure}

This decoupling leaves significant accuracy--cost gains unrealized because operator data formats jointly govern cross-layer numerical error on the software side and synthesized datapath structure on the hardware side.
First, as operator tensor distributions and workload shares vary significantly across networks and layers, quantization strategies are difficult to transfer directly across networks. As shown in Fig.~\ref{fig:motivation}(a), the relative accuracy rankings of different quantization strategies vary across networks. Therefore, a network-aware approach is required to adaptively select quantization schemes according to network-specific characteristics.
Second, MAC synthesis cost cannot be reliably captured by simple proxy models, because nominal bit width does not accurately reflect the true hardware cost of heterogeneous integer and custom floating-point MACs. As Fig.~\ref{fig:motivation}(b) shows, format pairs with identical bit-operation counts ($b_W b_A$) can exhibit nonlinear divergence in post-synthesis hardware costs due to structural differences.
Since the same heterogeneous formats yield divergent accuracy rankings across network topologies while exhibiting nonlinear area--delay trade-offs after logic synthesis, network-level format allocation and underlying MAC circuit synthesis must be co-optimized within a single closed-loop framework.

\begin{figure*}[!t]
\centering
\includegraphics[width=0.90\textwidth]{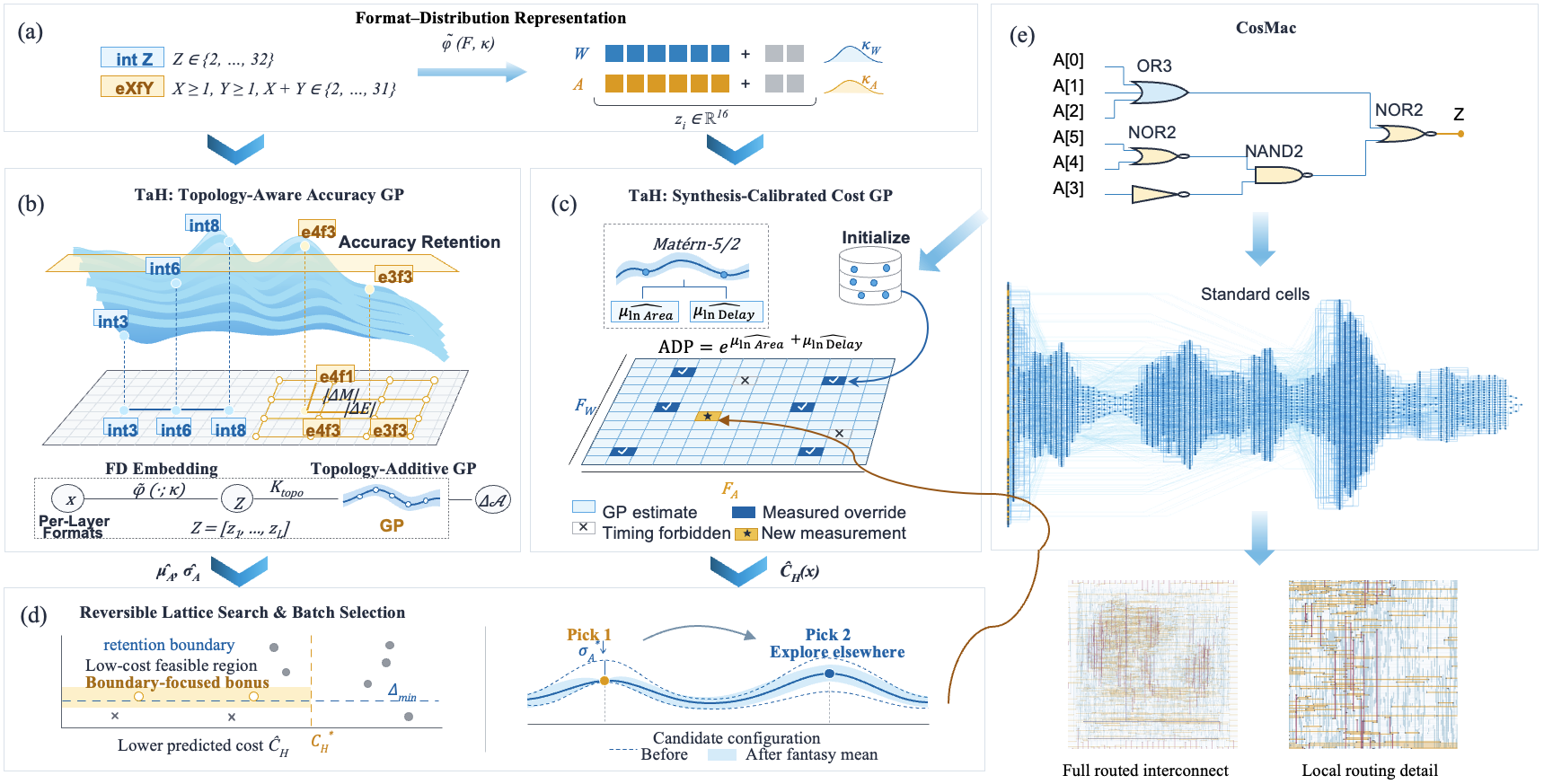}
\caption{The overview of NAMSI.}
\label{fig:method-overview}
\end{figure*}

Realizing such synthesis-in-the-loop customization under PTQ faces two fundamental challenges. 
\textbf{C1.} Full-network PTQ evaluation and gate-level MAC synthesis incur high runtime overhead, while cross-layer error coupling precludes decomposing the exponential design space into independent operator optimizations. \textbf{C2.} Heterogeneous formats admit no unified bit-width ordering across this discrete space, resulting in steep accuracy cliffs~\cite{kuzmin2022fp8} and a non-monotonic hardware-cost landscape.

To address these challenges, we make the following contributions:
\begin{itemize}
    \item We develop NAMSI, an end-to-end network-aware MAC synthesis framework that integrates parameterized heterogeneous MAC synthesis into full-network PTQ optimization under an accuracy constraint.
    \item We propose a topology-aware accuracy and hardware-cost model (TaH), pairing a computation-graph sparse Gaussian process (GP) with a synthesis-calibrated cost GP to jointly predict cross-layer coupled accuracy and post-synthesis ADP.
    \item We design a format--distribution representation (FD) that integrates format attributes with tensor statistics to characterize accuracy cliffs, and employ reversible lattice search (RLS) to efficiently explore the unordered and non-monotonic format space.
    \item We evaluate NAMSI across five ImageNet-1K backbones with $7\text{-nm}$ synthesis and placed-and-routed MAC implementations, achieving $54.2\%$--$60.4\%$ MAC ADP reductions over FP32 with $\ge 99\%$ accuracy retention under frozen PTQ.
\end{itemize}

\section{Preliminaries}
\label{sec:preliminaries}

\subsection{Data Formats}
\label{subsec:data-formats}

A numerical data format specifies how a finite sequence of bits represents a number. Its bit width determines the number of available encodings, while its encoding rules determine the range and spacing of representable values. Integer formats provide uniformly spaced values, whereas floating-point formats resemble scientific notation, using a sign, an exponent to set the magnitude, and a fraction to refine the value. At a fixed bit width, allocating more bits to the exponent expands the dynamic range, while allocating more bits to the fraction improves precision. 

\subsection{Problem Formulation}
\label{subsec:problem-formulation}

We formulate network-aware MAC synthesis as a constrained optimization problem
\begin{equation}
\begin{aligned}
\min_{x\in\mathcal{X}} \quad & C_H(x) \\
\mathrm{s.t.}\quad
& \tau\mathcal{A}_{\mathrm{FP}}-\mathcal{A}(x)\le 0, \\
& x\in\Omega_{\mathrm{num}}, \\
& x_i\in\Omega_{\mathrm{hw}},\quad i=1,\ldots,L.
\end{aligned}
\label{eq:opt-problem}
\end{equation}
Here, $x=(x_1,\ldots,x_L)$ specifies the weight--activation format pairs $x_i=(F_{W,i},F_{A,i})$ for $L$ MAC operators, with $\mathcal{X}=(\mathcal{F}^{2})^{L}$. The objective $C_H(x)=\sum_{i=1}^{L}w_i\,\mathrm{Area}(x_i)\mathrm{Delay}(x_i)$ is the workload-weighted post-synthesis MAC ADP, where $w_i$ is operator $i$'s share of the total MAC count. The accuracy constraint requires the quantized accuracy $\mathcal{A}(x)$ to retain at least a fraction $\tau$ of the full-precision accuracy $\mathcal{A}_{\mathrm{FP}}$ on the same evaluation split. The sets $\Omega_{\mathrm{num}}$ and $\Omega_{\mathrm{hw}}$ enforce numerical validity and timing closure under the target clock period, respectively.

\section{Method}
\label{sec:method}

\subsection{Framework Overview}
\label{subsec:overview}

NAMSI co-optimizes operator formats and MAC circuits in a closed loop (Fig.~\ref{fig:method-overview} and Algorithm~\ref{alg:master-framework}). It minimizes workload-weighted post-synthesis MAC ADP under accuracy, numerical-validity, and timing constraints. Offline initialization profiles per-tensor statistics on calibration data and extracts the MAC graph $G$. FD encodes each operand format together with its tensor distribution. TaH then predicts full-network accuracy and hardware cost. Next, RLS proposes candidate assignments and selects an evaluation batch $\mathcal{S}$. Before full-network PTQ, \cosmac{} synthesizes and verifies unseen format pairs in $\mathcal{S}$. New measurements update the surrogates, hardware cache $\mathcal{H}$, and search states $\Lambda$. Once the PTQ budget $B_{\mathrm{tot}}$ is exhausted, NAMSI returns the lowest-cost verified feasible assignment.

\begin{algorithm}[!t]
\caption{NAMSI Closed-Loop Format and MAC Synthesis}
\label{alg:master-framework}
\footnotesize
\begin{algorithmic}[1]
\REQUIRE Format domain $\mathcal F$, seed observations $\mathcal D_{\mathrm{init}}$, PTQ budget $B_{\mathrm{tot}}$, batch size $B$.
\ENSURE Best verified feasible assignment $x^*$.
\STATE Profile tensors and extract graph $G$.
\STATE Initialize hardware cache $\mathcal H$ and lattice states $\Lambda$.
\STATE $\mathcal D\leftarrow\mathcal D_{\mathrm{init}}$, $\mathcal U\leftarrow\{x:(x,\Delta\mathcal A)\in\mathcal D\}$, $n\leftarrow0$
\WHILE{$n<B_{\mathrm{tot}}$}
    \STATE $x^*\leftarrow\operatorname{BestFeasible}(\mathcal D,\mathcal H)$
    \STATE Fit accuracy GP on $\mathcal D$ and cost GPs on $\mathcal H$.
    \STATE $\mathcal P\leftarrow\operatorname{Propose}(\mathcal D,\Lambda,G,\mathcal H)\setminus\mathcal U$
    \IF{$\mathcal P=\varnothing$}
        \STATE \textbf{break}
    \ENDIF
    \STATE Select up to $\min\{B,B_{\mathrm{tot}}-n,|\mathcal P|\}$ candidates for $\mathcal S$ via \eqref{eq:cecr-acquisition} and Kriging Believer.
    \FOR{$x\in\mathcal S$}
        \STATE $\mathcal U\leftarrow\mathcal U\cup\{x\}$
        \STATE Verify unseen pairs in $x$ and update $\mathcal H$.
        \IF{all pairs pass hardware verification}
            \STATE $(\mathcal A(x),v)\leftarrow\operatorname{PTQ}(x)$
            \STATE $n\leftarrow n+1$
            \IF{$v=\mathrm{valid}$}
                \STATE $\mathcal D\leftarrow\mathcal D\cup\{(x,\mathcal A(x)-\mathcal A_{\mathrm{FP}})\}$
            \ENDIF
        \ENDIF
        \STATE Update $\Lambda$ using verification and PTQ outcomes.
    \ENDFOR
\ENDWHILE
\RETURN $\operatorname{BestFeasible}(\mathcal D,\mathcal H)$
\end{algorithmic}
\end{algorithm}

\subsection{Parameterized MAC Synthesis via \cosmac{}}
\label{subsec:cosmac-synthesis}

For each format pair $x_i=(F_{W,i},F_{A,i})$, \cosmac{} generates specialized SystemVerilog RTL for a fused MAC with FP32 accumulation. Format parameters configure operand decoding, multiplication, accumulator alignment, normalization, and rounding. For each pair, \cosmac{} evaluates multiple arithmetic structures and pipeline configurations, retaining the lowest-cost valid design. Synopsys Design Compiler synthesizes the RTL and checks timing against $T_{\mathrm{clk}}$. Candidates enter $\Omega_{\mathrm{hw}}$ only after passing gate-level functional verification and setup/hold timing checks.

The hardware cache $\mathcal{H}$ stores each evaluated pair's measured cost $c_{\mathrm{MAC}}(x_i)=\mathrm{Area}(x_i)\mathrm{Delay}(x_i)$ and verification status. Cached results are reused across operators and search rounds, so synthesis runs only for unseen pairs in $\mathcal{S}$.

\subsection{Format--Distribution Representation}
\label{subsec:format-embedding}

FD represents each format $F\in\mathcal{F}$ with six numerical attributes:
\begin{equation}
\phi(F)=\left[k(F),m_{\mathrm{eff}}(F),e(F),b(F),r(F),q(F)\right].
\label{eq:format-embedding}
\end{equation}
Here, $k\in\{0,1\}$ indicates integer or floating point. The term $m_{\mathrm{eff}}$ equals $Z-1$ for signed $\mathrm{int}Z$ and $Y$ for $\mathrm{e}X\mathrm{f}Y$. The attributes $e$ and $b$ denote exponent width and total bit width, giving $e=0,b=Z$ for $\mathrm{int}Z$ and $e=X,b=1+X+Y$ for $\mathrm{e}X\mathrm{f}Y$. Under standard exponent bias and subnormal support, $r=\log_2(\mathrm{max\_finite}/\mathrm{min\_pos})$ measures log dynamic range, while $q=-m_{\mathrm{eff}}$ measures log resolution.

Each tensor $T$ is summarized by its log crest factor:
\begin{equation}
\kappa(T)=\log_2\!\left(\frac{\lVert T\rVert_\infty+\varepsilon}{\sqrt{\mathbb E[T^2]}+\varepsilon}\right),
\label{eq:crest-factor}
\end{equation}
where $\varepsilon>0$ ensures numerical stability. Larger $\kappa(T)$ indicates stronger outliers relative to root-mean-square magnitude. Under peak-scaled quantization, outliers dilute uniform integer resolution across the tensor bulk according to $m_{\mathrm{eff}}(F)-\kappa(T)$. Narrow formats also incur a range-span deficit $\max(0,\kappa(T)-r(F))$, while $\kappa(T)k(F)$ captures floating-point tail adaptation. Bounded nonlinear scaling of $\phi(F)$ and these interaction terms yields an operand embedding $\tilde\phi(F,\kappa)\in\mathbb{R}^8$. Concatenating weight and activation embeddings gives
\begin{equation}
\begin{aligned}
z_i(x)={}&\tilde\phi(F_{W,i},\kappa(W_i))\mathbin{\|}\tilde\phi(F_{A,i},\kappa(A_i)),\\
Z(x)={}&z_1(x)\mathbin{\|}\cdots\mathbin{\|}z_L(x),
\end{aligned}
\label{eq:operator-embedding}
\end{equation}
where $\|$ denotes concatenation and $z_i(x)\in\mathbb{R}^{16}$. This representation separates dynamic range from bulk precision while preserving outlier statistics.

\subsection{Topology-Aware Accuracy and Hardware-Cost Modeling}
\label{subsec:topology-sparse-surrogate}

TaH pairs a format-pair hardware cost model with a graph-based accuracy model. Using $\mathcal{H}$, two GPs predict area and delay from $\phi(F_{W,i})\mathbin{\|}\phi(F_{A,i})$. For unseen pairs, the predicted cost is
\begin{equation}
\begin{aligned}
\ln \hat c_{\mathrm{MAC}}(x_i)
&= \hat\mu_{\ln\mathrm{Area}}(x_i)
 + \hat\mu_{\ln\mathrm{Delay}}(x_i),\\
\hat C_H(x)
&= \sum_{i=1}^{L} w_i\,\hat c_{\mathrm{MAC}}(x_i).
\end{aligned}
\label{eq:pair-cost-estimate}
\end{equation}
Pairs in $\mathcal{H}$ directly use $c_{\mathrm{MAC}}(x_i)$, whereas failing pairs are rejected before scoring. 

For accuracy, execution tracing builds a sparse undirected graph $G=(\mathcal{V},\mathcal{E})$ over MAC operators. Edges connect nearest MAC ancestors along data-flow paths and branches meeting at residual additions. Each PTQ evaluation calibrates quantization ranges on the calibration split before measuring accuracy $\mathcal{A}(x)$. Runs without overflow, non-finite values, or zero-range collapse are marked valid ($x\in\Omega_{\mathrm{num}}$). Trained on valid records in $\mathcal{D}$ with sample-size-scaled noise variance, the accuracy GP models relative change $\Delta\mathcal{A}(x)=\mathcal{A}(x)-\mathcal{A}_{\mathrm{FP}}$ against each split's FP32 baseline. Its covariance combines node and edge terms:
\begin{equation}
\begin{split}
K_{\mathrm{topo}}(x,x')={}&\frac{\sigma_V^2 g_V(Z,Z')}{L}
\sum_{i=1}^{L}k_V(z_i,z_i')\\
&+\frac{\sigma_E^2 g_E(Z,Z')}{|\mathcal E|}
\sum_{(i,j)\in\mathcal E}k_E(z_i\mathbin{\|}z_j,z_i'\mathbin{\|}z_j'),
\end{split}
\label{eq:topo-kernel}
\end{equation}
where $Z=Z(x)$, $Z'=Z(x')$, $z_i=z_i(x)$, and $z_i'=z_i(x')$. Parameters $\sigma_V^2$ and $\sigma_E^2$ control node and edge amplitudes, and the edge sum is omitted when $\mathcal{E}=\varnothing$. This structure captures local sensitivity and adjacent-layer coupling without dense all-pair terms. From the GP posterior, TaH reports predicted accuracy $\mu_A(x)=\mathcal{A}_{\mathrm{FP}}+\mathbb{E}[\Delta\mathcal{A}(x)\mid\mathcal{D}]$ and standard deviation $\sigma_A(x)$.

\subsection{Reversible Lattice Search and Batch Selection}
\label{subsec:lattice-and-proposals}

RLS explores $\mathcal{X}$ using within-family capacity orders. The state map $\Lambda$ tracks per-operand format states (\emph{active}, \emph{dormant}, or \emph{forbidden}) alongside globally forbidden timing-violating pairs. Single-operator numerical failures forbid the offending format and its lower lattice cone. Severe local accuracy drops or cost-dominated formats enter the dormant state. When a connected neighbor in $G$ changes format, RLS probabilistically probes dormant boundary formats and reactivates those passing surrogate screening and verification.

Maintaining a small elite pool of the lowest-cost feasible assignments, RLS generates a deduplicated candidate pool $\mathcal{P}$ outside $\mathcal{U}$ in each round. Operators are sampled with probability proportional to $\sqrt{w_i}$. Proposals combine four moves: within-family lattice steps, $\kappa$-guided integer--floating-point switches at comparable bit widths, coupled precision redistribution across edges in $\mathcal{E}$, and joint compression of high-workload operators.

Let $C_H^*$ be the lowest measured feasible cost so far. The predicted relative cost reduction is $I_{\mathrm{cost}}(x)=\max(0,1-\hat C_H(x)/C_H^*)$. The feasibility probability is $P_{\mathrm{feas}}(x)=\Phi((\mu_A(x)-\tau\mathcal A_{\mathrm{FP}})/\allowbreak\max(\sigma_A(x),\varepsilon))$, where $\Phi$ is the standard normal CDF. Each candidate is scored by
\begin{equation}
\begin{split}
\alpha(x)={}&I_{\mathrm{cost}}(x)P_{\mathrm{feas}}(x)
+\lambda\,\mathbb I[\hat C_H(x)<C_H^*]\\
&\times\frac{\sigma_A(x)}{\mathcal A_{\mathrm{FP}}}
\exp\!\left[-\frac{(\mu_A(x)-\tau\mathcal A_{\mathrm{FP}})^2}{2\sigma_A^2(x)+\varepsilon}\right],
\end{split}
\label{eq:cecr-acquisition}
\end{equation}
where $\mathbb{I}[\cdot]$ is the indicator function and $\lambda>0$ weights exploration near the accuracy boundary. Kriging Believer~\cite{ginsbourger2010kriging} greedily selects up to $\min\{B,B_{\mathrm{tot}}-n,|\mathcal{P}|\}$ candidates for batch $\mathcal{S}$. After each pick, a temporary accuracy GP is conditioned on that candidate's predicted mean $\mu_A(x)$ to suppress redundant picks. Here, $B$ is the maximum batch size and $n$ counts online PTQ runs. Each $x\in\mathcal{S}$ undergoes \cosmac{} verification before PTQ, and only PTQ runs consume $B_{\mathrm{tot}}$. Valid PTQ runs update $\mathcal{D}$ even when below $\tau\mathcal{A}_{\mathrm{FP}}$, while measured costs update $\mathcal{H}$. NAMSI returns the feasible assignment with the lowest measured cost.

\section{Experimental Evaluation}
\label{sec:experiments}

\subsection{Experimental Setup}
\label{subsec:exp-setup}

All experiments run in PyTorch~2.5.1 with CUDA~12.1 on Linux servers. For \cosmac{}-generated MAC RTL, logic synthesis and physical implementation are performed using Synopsys Design Compiler and Synopsys IC Compiler~II (ICC2), respectively, based on the $7\text{-nm}$ ASAP7 PDK~\cite{clark2016asap7} at $T_{\mathrm{clk}} = 2{,}000\text{ ps}$.

We evaluate five networks on ImageNet-1K~\cite{deng2009imagenet}: ResNet-18, ResNet-34~\cite{he2016resnet}, VGG16-BN~\cite{simonyan2015vgg}, RegNetX-400MF~\cite{radosavovic2020regnet}, and MobileNetV2~\cite{sandler2018mobilenetv2}. 
All weights remain frozen without fine-tuning, using FP32 Top-1 accuracy ($\mathcal{A}_{\mathrm{FP}}$) as the reference with a uniform retention threshold of $\tau = 0.99$. We sample two disjoint training subsets for calibration ($C$, $4{,}000$ images) and online search evaluation ($S$, $1{,}000$--$5{,}000$ images), keeping the official $50{,}000$-image validation set ($V$) strictly unseen until a single final evaluation.

We compare NAMSI against Qualcomm AIMET-AMP (v1.35.0)~\cite{siddegowda2022aimet}, which searches independent weight and activation formats in $\{\mathrm{int}2,\ldots,\mathrm{int}8\}$ under the identical frozen PTQ protocol and measures hardware ADP through the same \cosmac{} synthesis flow. We also set two ablation studies to respectively validate the effectiveness of surrogate-guided reversible lattice search against sequential per-layer greedy selection (Greedy-Sensitivity) and the necessity of cross-layer topological coupling by disabling graph edges (NAMSI w/o Edge). Across our experiments, search is initialized with $256$-image offline single-operator anchors and optimized over an online evaluation budget of $B_{\mathrm{tot}} = 100$.

\subsection{End-to-End Accuracy and Hardware Comparison}
\label{subsec:exp-main}

\begin{figure*}[!t]
    \centering
    \includegraphics[width=0.9\textwidth]{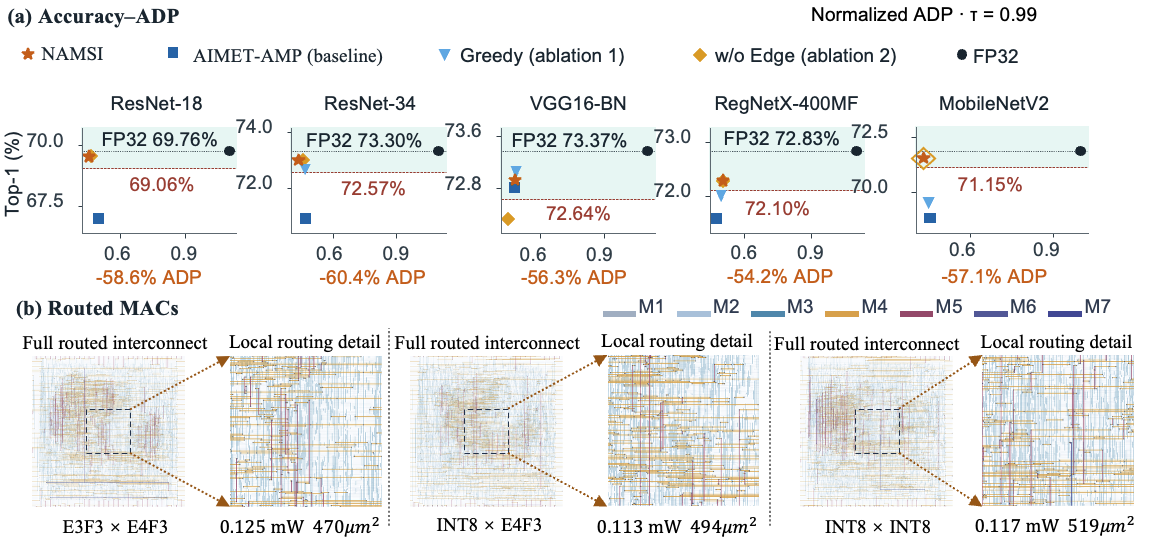}
    \caption{End-to-end comparison across networks and routed MAC layouts. (a) Locked validation ($V$) Top-1 accuracy versus FP32-normalized workload-weighted MAC ADP. (b) ICC2 placed-and-routed MAC layouts showing the full routed interconnect and local routing details.}
    \label{fig:exp-main}
\end{figure*}

Figure~\ref{fig:exp-main}(a) summarizes the locked $V$-split accuracy and post-synthesis hardware costs across networks, with hardware metrics normalized to the synthesized FP32 MAC cost for cross-network alignment. Across evaluated networks, NAMSI cuts MAC ADP by $\mathbf{54.2\%}$--$\mathbf{60.4\%}$ relative to FP32, achieving lower or comparable ADP against AIMET-AMP. While AIMET-AMP incurs substantial accuracy degradation, NAMSI consistently preserves at least $\mathbf{99\%}$ of the full-precision FP32 baseline accuracy across all tested networks. For the compact grouped-convolution network RegNetX-400MF, AIMET-AMP suffers a pronounced relative accuracy drop due to limited inter-channel redundancy. In contrast, NAMSI selectively assigns heterogeneous formats with wider dynamic ranges to bottleneck-sensitive layers, reducing ADP by $54.2\%$ relative to FP32 while keeping the relative accuracy loss within $0.75\%$ and remaining strictly feasible.

Figure~\ref{fig:exp-main}(b) presents the routed MAC implementations, physically demonstrating that heterogeneous datapaths effectively reduce multiplier-related silicon area overhead that is not captured by nominal bit-width proxies. Collectively, these end-to-end and post-route results confirm the effectiveness of NAMSI, demonstrating clear accuracy--hardware advantages over conventional mixed-precision allocation.

To explain the mechanisms underlying these end-to-end gains, we next analyze NAMSI's decisions through layer-wise heterogeneous format allocation and tensor-level format--distribution matching.

\subsubsection{Layer-wise Heterogeneous Format Allocation}
\label{subsec:exp-layerwise}

\begin{figure*}[!t]
    \centering
    \includegraphics[width=0.90\textwidth]{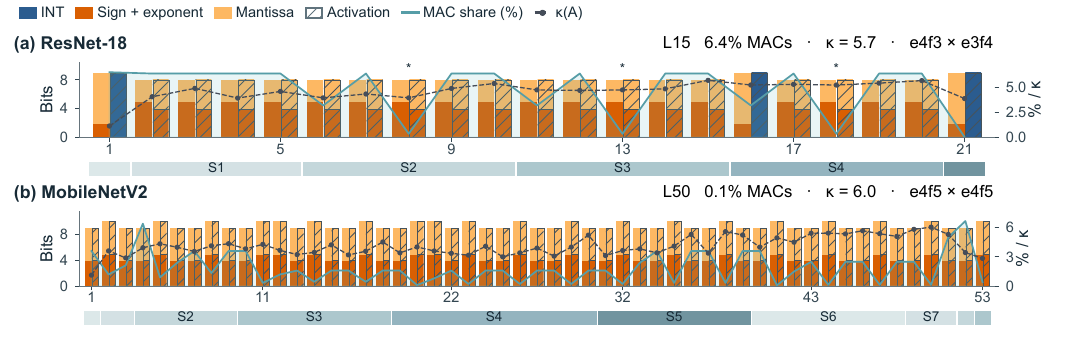}
    \caption{Layer-wise weight (solid) and activation (hatched) format allocations discovered by NAMSI, overlaid with per-layer MAC shares ($w_i$) and activation log crest factors ($\kappa(A_i)$). Floating-point bars decompose total width into $1+X+Y$, where $X$ and $Y$ are the exponent and fraction widths.}
    \label{fig:exp-layerwise}
\end{figure*}

The layer-wise heterogeneous format assignments identified across different networks reveal how NAMSI translates operator-level decisions into end-to-end accuracy--ADP gains. As shown in Fig.~\ref{fig:exp-layerwise}, rather than applying a uniform bit width or restricting all operators to a single format family, NAMSI assigns compact custom floating-point datapaths with wide dynamic ranges to compute-intensive layers exhibiting pronounced activation outliers, while retaining higher-precision formats for accuracy-critical bottlenecks.

For ResNet-18 (Fig.~\ref{fig:exp-layerwise}(a)), NAMSI allocates $9$-bit hybrid formats ($\mathrm{e}1\mathrm{f}7 \times \mathrm{int}9$) to three sensitive bottlenecks—the stem convolution with compact input activations ($\kappa(A_1) = 1.1$), the Stage-4 entry layer ($3.2\%$ MACs), and the final classifier ($0.03\%$ MACs)—preserving feature and output fidelity at bounded hardware cost. Conversely, across the remaining $18$ trunk and projection convolutions that dominate computation ($90.3\%$ of total MACs, up to $6.4\%$ per layer) and exhibit strong activation outliers ($\kappa(A_i) \in [3.9, 5.7]$, peaking at L15 with $\kappa = 5.7$), NAMSI compresses operands into compact $8$-bit custom floating-point pairs ($\mathrm{e}4\mathrm{f}3 \times \mathrm{e}3\mathrm{f}4$) to absorb heavy tails while maximizing network-wide ADP savings. 
These results demonstrate that effective heterogeneous format allocation must jointly account for each operator's workload contribution and tensor outlier statistics. NAMSI exploits both signals to adapt the format family and precision of each layer, balancing local accuracy sensitivity against network-wide hardware efficiency.

\subsubsection{Format--Distribution Mechanism}
\label{subsec:exp-micro}

\begin{figure*}[!t]
    \centering
    \includegraphics[width=0.90\textwidth]{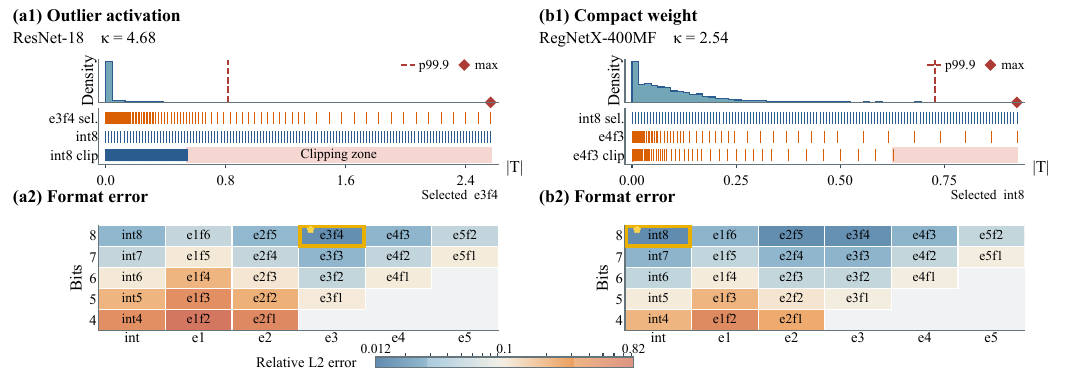}
    \caption{Microscopic format--distribution matching on two representative tensors. (a1)--(a2) Outlier-heavy activation in ResNet-18 ($\kappa = 4.68$), comparing representable quantization grids and $2$D relative $L_2$ error across $4$--$8$ bits. (b1)--(b2) Compact weight tensor in RegNetX-400MF ($\kappa = 2.54$).}
    \label{fig:exp-micro}
\end{figure*}

Figure~\ref{fig:exp-micro} contrasts an outlier-heavy ResNet-18 activation ($\kappa = 4.68$) with a compact RegNetX-400MF weight tensor ($\kappa = 2.54$) to expose the numerical mechanism behind these format choices across different tensor regimes. It shows how integer and floating-point quantization grids align with each magnitude distribution (top) and how the resulting relative $L_2$ quantization error varies across $4$--$8$-bit formats (bottom).
For ResNet-18 (Fig.~\ref{fig:exp-micro}(a1)), most values concentrate tightly near zero. Full-range $\mathrm{int}8$ inefficiently allocates uniform quantization levels to the sparsely populated tail, whereas percentile-clipped $\mathrm{int}8$ incurs severe error by truncating extreme values in the clipping zone. In contrast, the selected $\mathrm{e}3\mathrm{f}4$ format uses three exponent bits to cover the long tail and four mantissa bits to provide finer resolution near zero, thereby achieving the lowest relative $L_2$ error among the evaluated $8$-bit formats (Fig.~\ref{fig:exp-micro}(a2)).

For the compact RegNetX-400MF weight tensor (Fig.~\ref{fig:exp-micro}(b1)), magnitudes decay smoothly without pronounced outliers. Formats with wider exponent fields sacrifice effective precision and leave coarser quantization gaps across the densely populated mid-range, whereas uniform $\mathrm{int}8$ densely covers the tensor support and achieves the lowest error (Fig.~\ref{fig:exp-micro}(b2)). 
Overall, these results provide a microscopic numerical explanation for NAMSI's operator-level format choices. By matching the format family and exponent--mantissa allocation to the outlier characteristics of each tensor, NAMSI avoids clipping errors in heavy-tailed distributions while preserving resolution in compact distributions without allocating unnecessary exponent bits.

\subsection{Ablation Study}
\label{subsec:exp-ablation}

To isolate the contributions of RLS and TaH, we compare NAMSI against Greedy-Sensitivity and NAMSI w/o Edge under the identical evaluation budget and locked $V$-split protocol.

\subsubsection{Ablation 1:}
Greedy-Sensitivity compresses layers sequentially based on isolated single-operator sensitivity. As shown in Fig.~\ref{fig:exp-main}(a), this myopic strategy violates the $\tau = 0.99$ accuracy constraint on compact networks, with MobileNetV2 accuracy \(2.00\) pp below NAMSI, while also yielding a $7.1\%$ higher MAC ADP on deeper backbones such as ResNet-34. 
These results demonstrate the importance of global surrogate-guided RLS in coordinating format choices across layers under stringent accuracy constraints.

\subsubsection{Ablation 2:}
Disabling graph edges in NAMSI w/o Edge ($\sigma_E = 0$ in \eqref{eq:topo-kernel}) ignores cross-layer error coupling. 
Without modeling sequential error accumulation, NAMSI w/o Edge over-compresses VGG16-BN and violates the $\tau = 0.99$ accuracy-retention threshold, as shown in Fig.~\ref{fig:exp-main}(a). Its failure to capture cross-branch interactions also results in $2.0\%$ and $4.9\%$ higher MAC ADP on ResNet-18 and ResNet-34, respectively. These results demonstrate the importance of TaH in capturing inter-layer error propagation and reducing hardware cost while maintaining feasibility.

\subsection{Search Efficiency and Runtime Breakdown}
\label{subsec:exp-runtime}

\begin{figure}[!t]
    \centering
    \includegraphics[width=0.85\columnwidth]{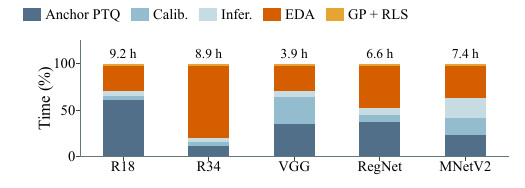}
    \caption{End-to-end search time and normalized five-stage runtime composition across evaluated networks.}
    \label{fig:exp-runtime}
\end{figure}

In each search round, RLS generates $\sim\!600$ candidate configurations containing many previously unsynthesized format pairs. Rather than invoking gate-level synthesis during proposal scoring, NAMSI evaluates unseen pairs via our cost GP and scores each $L$-layer candidate in milliseconds with $O(L)$ weighted summation \eqref{eq:pair-cost-estimate}. Exact synthesis is triggered only for newly encountered format pairs within the selected acquisition batch, with measured results cached in $\mathcal{H}$ for immediate reuse across layers and subsequent rounds. This two-speed evaluation significantly reduces the computational overhead associated with synthesis-in-the-loop optimization.

Figure~\ref{fig:exp-runtime} further breaks down end-to-end runtime into five stages: offline single-operator anchor PTQ (\texttt{Anchor PTQ}), online PTQ calibration (\texttt{Calib.}) and search-split inference (\texttt{Infer.}), on-demand Design Compiler synthesis (\texttt{EDA}), and topology-aware GP fitting with RLS acquisition (\texttt{GP + RLS}). Across all evaluated networks, \texttt{GP + RLS} accounts for $\mathbf{< 3\%}$ of total runtime. The remaining runtime is instead spent primarily on essential PTQ evaluation and exact synthesis of newly encountered heterogeneous format pairs.
Notably, ResNet-34 exhibits a markedly higher EDA runtime fraction because its search encounters more previously unseen format pairs, resulting in a lower synthesis-cache hit rate and more frequent on-demand Design Compiler invocations.

\section{Conclusion}
\label{sec:conclusion}

We presented NAMSI, a network-aware automated MAC synthesis framework that co-optimizes operator-level heterogeneous data formats and \cosmac{}-synthesized arithmetic datapaths under PTQ. By integrating our format--distribution representation, dual-GP topology-aware accuracy and hardware-cost modeling, and reversible lattice search, NAMSI navigates the unordered format space while invoking gate-level synthesis only on unseen format pairs. Across five networks on ImageNet-1K in $7\text{-nm}$ ASAP7 technology, NAMSI cuts post-synthesis MAC ADP by $54.2\%$--$60.4\%$ over FP32 with $\ge 99\%$ Top-1 accuracy retention within $3.9$--$9.2\text{ h}$, demonstrating the importance of jointly optimizing the network and its hardware implementation.

\bibliographystyle{IEEEtran}
\bibliography{references}

@inproceedings{he2016resnet,
  author = {He, Kaiming and Zhang, Xiangyu and Ren, Shaoqing and Sun, Jian},
  title = {Deep Residual Learning for Image Recognition},
  booktitle = {Proceedings of the IEEE Conference on Computer Vision and Pattern Recognition (CVPR)},
  year = {2016},
  pages = {770--778},
}

@inproceedings{wang2019haq,
  author = {Wang, Kuan and Liu, Zhijian and Lin, Yujun and Lin, Ji and Han, Song},
  title = {{HAQ}: Hardware-Aware Automated Quantization With Mixed Precision},
  booktitle = {Proceedings of the IEEE/CVF Conference on Computer Vision and Pattern Recognition (CVPR)},
  year = {2019},
  pages = {8604--8612},
}

@inproceedings{yao2021hawqv3,
  author = {Yao, Zhewei and Dong, Zhen and Zheng, Zhangcheng and Gholami, Amir and Yu, Jiali and Tan, Eric and Wang, Leyuan and Huang, Qijing and Wang, Yida and Mahoney, Michael W. and Keutzer, Kurt},
  title = {{HAWQ-V3}: Dyadic Neural Network Quantization},
  booktitle = {Proceedings of the 38th International Conference on Machine Learning (ICML)},
  year = {2021},
  volume = {139},
  pages = {11875--11886},
}

@inproceedings{cai2020edmips,
  author = {Cai, Zhaowei and Vasconcelos, Nuno},
  title = {Rethinking Differentiable Search for Mixed-Precision Neural Networks},
  booktitle = {Proceedings of the IEEE/CVF Conference on Computer Vision and Pattern Recognition (CVPR)},
  year = {2020},
  pages = {2349--2358},
}

@inproceedings{nagel2020adaround,
  author = {Nagel, Markus and Amjad, Rana Ali and Van Baalen, Mart and Louizos, Christos and Blankevoort, Tijmen},
  title = {Up or Down? {Adaptive} Rounding for Post-Training Quantization},
  booktitle = {Proceedings of the 37th International Conference on Machine Learning (ICML)},
  year = {2020},
  volume = {119},
  pages = {7197--7206},
}

@inproceedings{li2021brecq,
  author = {Li, Yuhang and Gong, Ruihao and Tan, Xu and Yang, Yang and Hu, Peng and Zhang, Qi and Yu, Fengwei and Wang, Wei and Gu, Shi},
  title = {{BRECQ}: Pushing the Limit of Post-Training Quantization by Block Reconstruction},
  booktitle = {International Conference on Learning Representations (ICLR)},
  year = {2021},
}

@inproceedings{sharma2018bitfusion,
  author = {Sharma, Hardik and Park, Jongse and Suda, Naveen and Lai, Liangzhen and Chau, Benson and Kim, Joon Kyung and Chandra, Vikas and Esmaeilzadeh, Hadi},
  title = {Bit Fusion: Bit-Level Dynamically Composable Architecture for Accelerating Deep Neural Networks},
  booktitle = {Proceedings of the ACM/IEEE 45th Annual International Symposium on Computer Architecture (ISCA)},
  year = {2018},
  pages = {764--775},
}

@inproceedings{jacob2018quantization,
  author = {Jacob, Benoit and Kligys, Skirmantas and Chen, Bo and Zhu, Menglong and Tang, Matthew and Howard, Andrew and Adam, Hartwig and Kalenichenko, Dmitry},
  title = {Quantization and Training of Neural Networks for Efficient Integer-Arithmetic-Only Inference},
  booktitle = {Proceedings of the IEEE Conference on Computer Vision and Pattern Recognition (CVPR)},
  year = {2018},
  pages = {2704--2713},
}

@inproceedings{judd2016stripes,
  author = {Judd, Patrick and Albericio, Jorge and Hetherington, Tayler and Aamodt, Tor M. and Moshovos, Andreas},
  title = {{Stripes}: Bit-Serial Deep Neural Network Computing},
  booktitle = {Proceedings of the 49th Annual IEEE/ACM International Symposium on Microarchitecture (MICRO)},
  year = {2016},
  pages = {1--12},
}

@inproceedings{tambe2020adaptivfloat,
  author = {Tambe, Thierry and Yang, En-Yu and Wan, Zishen and Deng, Yuntian and Reddi, Vijay Janapa and Rush, Alexander and Brooks, David and Wei, Gu-Yeon},
  title = {Algorithm-Hardware Co-Design of Adaptive Floating-Point Encodings for Resilient Deep Learning Inference},
  booktitle = {Proceedings of the 57th ACM/IEEE Design Automation Conference (DAC)},
  year = {2020},
  pages = {1--6},
}

@inproceedings{rouhani2020msfp,
  author = {Darvish Rouhani, Bita and Lo, Daniel and Zhao, Ritchie and Liu, Ming and Fowers, Jeremy and Ovtcharov, Kalin and Vinogradsky, Anna and Massengill, Sarah and Yang, Lita and Bittner, Ray and Forin, Alessandro and Zhu, Haishan and Na, Taesik and Patel, Prerak and Che, Shuai and Koppaka, Lok Chand and Song, Xia and Som, Subhojit and Das, Kaustav and Tiwary, Saurabh and Reinhardt, Steve and Lanka, Sitaram and Chung, Eric and Burger, Doug},
  title = {Pushing the Limits of Narrow Precision Inferencing at Cloud Scale with {Microsoft} Floating Point},
  booktitle = {Advances in Neural Information Processing Systems (NeurIPS)},
  year = {2020},
  volume = {33},
  pages = {10271--10281},
}

@inproceedings{kuzmin2022fp8,
  author = {Kuzmin, Andrey and Van Baalen, Mart and Ren, Yuwei and Nagel, Markus and Peters, Jorn and Blankevoort, Tijmen},
  title = {{FP8} Quantization: The Power of the Exponent},
  booktitle = {Advances in Neural Information Processing Systems (NeurIPS)},
  year = {2022},
  volume = {35},
  pages = {14651--14662},
}

@inproceedings{crafton2025pareto,
  author = {Crafton, Brian and Peng, Xiaochen and Sun, Xiaoyu and Lele, Ashwin Sanjay and Zhang, Bo and Khwa, Win-San and Akarvardar, Kerem},
  title = {Finding the {Pareto} Frontier of Low-Precision Data Formats and {MAC} Architecture for {LLM} Inference},
  booktitle = {Proceedings of the 62nd ACM/IEEE Design Automation Conference (DAC)},
  year = {2025},
  pages = {1--7},
  doi = {10.1109/dac63849.2025.11132989},
}

@incollection{ginsbourger2010kriging,
  author = {Ginsbourger, David and Le Riche, Rodolphe and Carraro, Laurent},
  editor = {Tenne, Yoel and Goh, Chi-Keong},
  title = {Kriging Is Well-Suited to Parallelize Optimization},
  booktitle = {Computational Intelligence in Expensive Optimization Problems},
  year = {2010},
  pages = {131--162},
  publisher = {Springer},
  series = {Adaptation, Learning, and Optimization},
  doi = {10.1007/978-3-642-10701-6_6},
}

@inproceedings{rouhani2023microexponents,
  author = {Darvish Rouhani, Bita and Zhao, Ritchie and Elango, Venmugil and Shafipour, Rasoul and Hall, Mathew and Mesmakhosroshahi, Maral and More, Ankit and Melnick, Levi and Golub, Maximilian and Varatkar, Girish and Shao, Lai and Kolhe, Gaurav and Melts, Dimitry and Klar, Jasmine and L'Heureux, Renee and Perry, Matt and Burger, Doug and Chung, Eric and Deng, Zhaoxia (Summer) and Naghshineh, Sam and Park, Jongsoo and Naumov, Maxim},
  title = {With Shared Microexponents, A Little Shifting Goes a Long Way},
  booktitle = {Proceedings of the 50th Annual International Symposium on Computer Architecture (ISCA)},
  year = {2023},
  doi = {10.1145/3579371.3589351},
}

@inproceedings{liu2023llmfp4,
  author = {Liu, Shih-yang and Liu, Zechun and Huang, Xijie and Dong, Pingcheng and Cheng, Kwang-Ting},
  title = {{LLM-FP4}: 4-Bit Floating-Point Quantized Transformers},
  booktitle = {Proceedings of the 2023 Conference on Empirical Methods in Natural Language Processing (EMNLP)},
  year = {2023},
  pages = {592--605},
  doi = {10.18653/v1/2023.emnlp-main.39},
}

@inproceedings{li2020apot,
  author = {Li, Yuhang and Dong, Xin and Wang, Wei},
  title = {Additive Powers-of-Two Quantization: An Efficient Non-uniform Discretization for Neural Networks},
  booktitle = {International Conference on Learning Representations (ICLR)},
  year = {2020},
}

@inproceedings{dotzel2024sf4,
  author = {Dotzel, Jordan and Chen, Yuzong and Kotb, Bahaa and Prasad, Sushma and Wu, Gang and Li, Sheng and Abdelfattah, Mohamed S. and Zhang, Zhiru},
  title = {Learning from Students: Applying t-Distributions to Explore Accurate and Efficient Formats for {LLM}s},
  booktitle = {Proceedings of the 41st International Conference on Machine Learning (ICML)},
  year = {2024},
  volume = {235},
  pages = {11573--11591},
  series = {Proceedings of Machine Learning Research},
}

@inproceedings{esser2020lsq,
  author = {Esser, Steven K. and McKinstry, Jeffrey L. and Bablani, Deepika and Appuswamy, Rathinakumar and Modha, Dharmendra S.},
  title = {Learned Step Size Quantization},
  booktitle = {International Conference on Learning Representations (ICLR)},
  year = {2020},
}

@inproceedings{frantar2023gptq,
  author = {Frantar, Elias and Ashkboos, Saleh and Hoefler, Torsten and Alistarh, Dan},
  title = {{OPTQ}: Accurate Post-Training Quantization for Generative Pre-trained Transformers},
  booktitle = {International Conference on Learning Representations (ICLR)},
  year = {2023},
}

@inproceedings{xiao2023smoothquant,
  author = {Xiao, Guangxuan and Lin, Ji and Seznec, Mickael and Wu, Hao and Demouth, Julien and Han, Song},
  title = {{SmoothQuant}: Accurate and Efficient Post-Training Quantization for Large Language Models},
  booktitle = {Proceedings of the 40th International Conference on Machine Learning (ICML)},
  year = {2023},
  volume = {202},
  pages = {38087--38099},
  series = {Proceedings of Machine Learning Research},
}

@inproceedings{lin2024awq,
  author = {Lin, Ji and Tang, Jiaming and Tang, Haotian and Yang, Shang and Chen, Wei-Ming and Wang, Wei-Chen and Xiao, Guangxuan and Dang, Xingyu and Gan, Chuang and Han, Song},
  title = {{AWQ}: Activation-aware Weight Quantization for On-Device {LLM} Compression and Acceleration},
  booktitle = {Proceedings of Machine Learning and Systems (MLSys)},
  year = {2024},
  volume = {6},
}

@inproceedings{xia2024quantllm,
  author = {Xia, Haojun and Zheng, Zhen and Wu, Xiaoxia and Chen, Shiyang and Yao, Zhewei and Youn, Stephen and Bakhtiari, Arash and Wyatt, Michael and Zhuang, Donglin and Zhou, Zhongzhu and Ruwase, Olatunji and He, Yuxiong and Song, Shuaiwen Leon},
  title = {{Quant-LLM}: Accelerating the Serving of Large Language Models via {FP6}-Centric Algorithm-System Co-Design on Modern {GPUs}},
  booktitle = {Proceedings of the 2024 USENIX Annual Technical Conference (USENIX ATC)},
  year = {2024},
  pages = {699--713},
}

@inproceedings{dotzel2024fliqs,
  author = {Dotzel, Jordan and Wu, Gang and Li, Andrew and Umar, Muhammad and Ni, Yun and Abdelfattah, Mohamed S. and Zhang, Zhiru and Cheng, Liqun and Dixon, Martin G. and Jouppi, Norman P. and Le, Quoc V. and Li, Sheng},
  title = {{FLIQS}: One-Shot Mixed-Precision Floating-Point and Integer Quantization Search},
  booktitle = {Proceedings of the Third International Conference on Automated Machine Learning},
  year = {2024},
  volume = {256},
  pages = {6/1--26},
  series = {Proceedings of Machine Learning Research},
}

@inproceedings{zhang2024mofq,
  author = {Zhang, Yijia and Zhao, Lingran and Cao, Shijie and Zhang, Sicheng and Wang, Wenqiang and Cao, Ting and Yang, Fan and Yang, Mao and Zhang, Shanghang and Xu, Ningyi},
  title = {Integer or Floating Point? New Outlooks for Low-Bit Quantization on Large Language Models},
  booktitle = {Proceedings of the IEEE International Conference on Multimedia and Expo (ICME)},
  year = {2024},
  pages = {1--6},
  doi = {10.1109/icme57554.2024.10688089},
}

@inproceedings{guo2022ant,
  author = {Guo, Cong and Zhang, Chen and Leng, Jingwen and Liu, Zihan and Yang, Fan and Liu, Yunxin and Guo, Minyi and Zhu, Yuhao},
  title = {{ANT}: Exploiting Adaptive Numerical Data Type for Low-Bit Deep Neural Network Quantization},
  booktitle = {Proceedings of the 55th IEEE/ACM International Symposium on Microarchitecture (MICRO)},
  year = {2022},
  pages = {1414--1433},
  doi = {10.1109/micro56248.2022.00095},
}

@inproceedings{jang2025blockdialect,
  author = {Jang, Wonsuk and Tambe, Thierry},
  title = {{BlockDialect}: Block-Wise Fine-Grained Mixed Format Quantization for Energy-Efficient {LLM} Inference},
  booktitle = {Proceedings of the 42nd International Conference on Machine Learning (ICML)},
  year = {2025},
  volume = {267},
  pages = {26925--26945},
  series = {Proceedings of Machine Learning Research},
}

@inproceedings{deng2009imagenet,
  author = {Deng, Jia and Dong, Wei and Socher, Richard and Li, Li-Jia and Li, Kai and Fei-Fei, Li},
  title = {{ImageNet}: A Large-Scale Hierarchical Image Database},
  booktitle = {Proceedings of the IEEE Conference on Computer Vision and Pattern Recognition (CVPR)},
  year = {2009},
  pages = {248--255},
}

@inproceedings{simonyan2015vgg,
  author = {Simonyan, Karen and Zisserman, Andrew},
  title = {Very Deep Convolutional Networks for Large-Scale Image Recognition},
  booktitle = {International Conference on Learning Representations (ICLR)},
  year = {2015},
}

@inproceedings{sandler2018mobilenetv2,
  author = {Sandler, Mark and Howard, Andrew and Zhu, Menglong and Zhmoginov, Andrey and Chen, Liang-Chieh},
  title = {{MobileNetV2}: Inverted Residuals and Linear Bottlenecks},
  booktitle = {Proceedings of the IEEE/CVF Conference on Computer Vision and Pattern Recognition (CVPR)},
  year = {2018},
  pages = {4510--4520},
}

@inproceedings{radosavovic2020regnet,
  author = {Radosavovic, Ilija and Kosaraju, Raj Prateek and Girshick, Ross and He, Kaiming and Doll{\'a}r, Piotr},
  title = {Designing Network Design Spaces},
  booktitle = {Proceedings of the IEEE/CVF Conference on Computer Vision and Pattern Recognition (CVPR)},
  year = {2020},
  pages = {10428--10436},
}

@article{clark2016asap7,
  author = {Clark, Lawrence T. and Vashishtha, Vinay and Shifren, Lucian and Gujja, Aditya and Sinha, Saurabh and Cline, Brian and Ramamurthy, Chandarasekaran and Yeric, Greg},
  title = {{ASAP7}: A 7-nm FinFET Predictive Process Design Kit},
  journal = {Microelectronics Journal},
  year = {2016},
  volume = {53},
  pages = {105--115},
}

@article{siddegowda2022aimet,
  author = {Siddegowda, Sangeetha and Fournarakis, Marios and Nagel, Markus and Blankevoort, Tijmen and Patel, Chirag and Khobare, Abhijit},
  title = {Neural Network Quantization with {AI} Model Efficiency Toolkit ({AIMET})},
  journal = {arXiv preprint arXiv:2201.08442},
  year = {2022},
}

\end{document}